\documentclass[%
reprint,
superscriptaddress,
nofootinbib,
 amsmath,amssymb,
 prc,
]{revtex4-2}

\usepackage{graphicx}
\usepackage{dcolumn}
\usepackage{bm}
\usepackage{supertabular}
\usepackage{longtable}
\usepackage{color}
\usepackage{url}
\usepackage[toc,page]{appendix}
\usepackage{flushend}

\begin{document}


\title{Microscopic core-quasiparticle coupling model for spectroscopy of odd-mass nuclei with octupole correlations}


\author{W. Sun}
\affiliation{School of Physical Science and Technology, Southwest University, Chongqing 400715, China}
\author{S. Quan}
\affiliation{School of Physical Science and Technology, Southwest University, Chongqing 400715, China}%
\author{Z. P. Li}\email{zpliphy@swu.edu.cn}
\affiliation{School of Physical Science and Technology, Southwest University, Chongqing 400715, China}%
\affiliation{Department of Physics and Electronic Science, Qiannan Normal University for Nationalities, Duyun, 558000, China}

\author{J. Zhao}
\affiliation{Microsystem and Terahertz Research Center and Insititute of Electronic Engineering,
	China Academy of Engineering Physics, Chengdu 610200, Sichuan, China}
	
\author{T. Nik\v{s}i\'c}
\affiliation{Physics Department, Faculty of Science, University of Zagreb, Bijeni\v{c}ka Cesta 32, Zagreb 10000, Croatia}
\author{D. Vretenar}
\affiliation{Physics Department, Faculty of Science, University of Zagreb, Bijeni\v{c}ka Cesta 32, Zagreb 10000, Croatia}

\date{October 24, 2019}

\begin{abstract}
\begin{description}
\item[Background] Predictions of spectroscopic properties of low-lying states are critical for nuclear structure studies. Theoretical methods can be particularly involved for odd-mass nuclei because of the interplay between the unpaired nucleon and collective degrees of freedom. Only a few models have been developed for systems in which octupole collective degrees of freedom play a role.

\item[Purpose] We aim to predict spectroscopic properties of odd-mass nuclei characterized by octupole shape deformation, employing a model that describes single-particle and collective degrees of freedom within the same microscopic framework.

\item[Method] A microscopic core-quasiparticle coupling (CQC) model based on the covariant density functional theory is developed, which includes collective excitations of even-mass core nuclei and single-particle states of the odd nucleon, calculated using a quadrupole-octupole collective Hamiltonian combined with a constrained reflection-asymmetric relativistic Hartree-Bogoliubov model.

\item[Results] Model predictions for low-energy excitation spectra and transition rates of odd-mass radium isotopes $^{223, 225, 227}$Ra are shown to be in good agreement with available data.

\item[Conclusions] A microscopic CQC model based on covariant density functional theory has been developed for odd-mass nuclei
characterized by both quadrupole and octupole shape deformations. Theoretical results reproduce data in odd-mass Ra isotopes and provide useful predictions for future studies of octupole correlations in nuclei and related phenomena.

\end{description}
\end{abstract}

\maketitle



\section{\label{secI}Introduction}

Studies of reflection-asymmetric octupole shapes have been a recurrent theme of interest in nuclear structure physics over several decades. More recently, experimental evidence for stable octupole deformation in nuclei has been reported, e.g., in $^{224}$Ra \cite{Gaffney2013Nature199}, $^{144}$Ba \cite{Bucher2016Phys.Rev.Lett.112503}, and $^{146}$Ba \cite{Bucher2017Phys.Rev.Lett.152504}. This type of shape deformation occurs predominantly in nuclei with neutron (proton) number $N (Z) \approx 34$, $56$, $88$, and $134$, characterized by the presence of low-lying negative-parity bands as well as pronounced electric octupole transitions \cite{Butler1996Rev.Mod.Phys.349,Butler2016J.Phys.G.73002}.

Octupole deformation in nuclei with an odd nucleon number has attracted specific interest in connection with observed parity-doublet structures and enhanced Schiff moments \cite{Butler1996Rev.Mod.Phys.349,ENGEL2013PPNP21,Chupp2019RevModPhys.91.015001,Dobaczewski2018PRL232501}. In the last decade considerable effort has been devoted to studies of excited states in odd-mass nuclei characterized by static and dynamic octupole deformation  \cite{Lic2017JPG054002,Wisniewski2017PRC064301,Wang2017EPJA234,Ruchowska2015Phys.Rev.C34328,Li2014Phys.Rev.C47303,Rzaca-Urban2013Phys.Rev.C31305,Rzaca-Urban2012Phys.Rev.C44324,ChenXC2016PRC021301,Ding2017PRC024301,Ahmad2015Phys.Rev.C24313,ReviolW2014,Tandel2013Phys.Rev.C34319,ReviolW2009}. New levels, transitions, and/or dipole moments  have been measured for nuclei in different mass regions, e.g., neutron-rich nuclei around $^{144}$Ba \cite{Lic2017JPG054002,Wisniewski2017PRC064301,Wang2017EPJA234,Ruchowska2015Phys.Rev.C34328,Li2014Phys.Rev.C47303,Rzaca-Urban2013Phys.Rev.C31305,Rzaca-Urban2012Phys.Rev.C44324}, neutron-deficient nuclei approaching  $N, Z\approx 56$ \cite{Ding2017PRC024301,ChenXC2016PRC021301}, heavy nuclei around $^{224}$Ra \cite{Ahmad2015Phys.Rev.C24313,ReviolW2014,Tandel2013Phys.Rev.C34319,ReviolW2009}. Furthermore, it is expected that heavy odd-mass octupole-deformed nuclei, such as $^{225}$Ra, amplify atomic electric dipole moments (EDMs) and related nuclear Schiff moments that arise from time-reversal and parity-violating interactions in the nuclear medium \cite{Bishof2016,ParkerRH2015,Dobaczewski2005PRL232502,Auerbach1996PRL4316,Flambaum2019PRC035501}. Effects of symmetry violation are greatly enhanced in $^{225}$Ra by collective effects and a closely spaced parity-doublet structure resulting from octupole nuclear deformation \cite{ParkerRH2015,Auerbach1996PRL4316}. As a wealth of new data on odd-A systems become accessible, it is important to develop consistent and accurate models that can be employed in theoretical studies of their spectroscopic properties.

A variety of theoretical methods have been applied to studies of octupole nuclear shapes and the evolution of the corresponding negative-parity collective states: self-consistent mean-field approaches \cite{Afanasjev2018Phys.Scr.34002,Fu2018Phys.Rev.C24338,Marevic2018Phys.Rev.C24334,Nomura2018Phys.Rev.C24317,Zhang2017Phys.Rev.C54308,Agbemava2017Phys.Rev.C24301,Yao2016Phys.Rev.C14306,Geng2007Chin.Phys.Lett.1865,Guo2010Phys.Rev.C47301,Robledo2010Phys.Rev.C34315,Robledo2011Phys.Rev.C54302,Rodriguez-Guzman2012Phys.Rev.C34336,Robledo2013Phys.Rev.C51302,Bernard2016Phys.Rev.C61302,Robledo2015J.Phys.G55109,Zhao2012Phys.Rev.C57304,Zhou2016Phys.Scr.63008,Nomura2013Phys.Rev.C21303,Nomura2014Phys.Rev.C24312,Nomura2015Phys.Rev.C14312,Agbemava2016Phys.Rev.C44304,Yao2016Phys.Rev.C11303,Zhang2010Phys.Rev.C34302,Li2013Phys.Lett.B866,Li2016J.Phys.G24005,Xia2017Phys.Rev.C54303,Yao2015Phys.Rev.C41304,Zhou2016Phys.Lett.B227,Tao2017Phys.Rev.C24319,Ebata2017Phys.Scr.64005}, macroscopic+microscopic (MM) models \cite{Moeller2008Atom.DataNucl.DataTabl.758,Wang2015Phys.Rev.C24303,Nazarewicz1984Nucl.Phys.A269}, algebraic (or interacting boson) models \cite{Zamfir2001PhysRevC054306,Otsuka1988Phys.Lett.B140}, phenomenological collective models \cite{Dobrowolski2016Phys.Rev.C54322,Minkov2012Phys.Rev.C34306,Jolos2012Phys.Rev.C24319,Bizzeti2013Phys.Rev.C11305,Bonatsos2005PhysRevC064309,Bizzeti2004PhysRevC064319}, and the reflection asymmetric shell model \cite{Chen2000PhysRevC.63.014314}.
For the case of odd-A nuclei theoretical approaches include the particle-rotor model \cite{LEANDER1984NPA375,LEANDER1985PLB284,Leander1987PRC1145,Leander1988PRC2744,Minkov2007PRC034324,Minkov2017PRL212501,WANG2019PLB454}, the cluster model \cite{Adamian2004}, the reflection-asymmetric shell model \cite{Chen2015Phys.Rev.C14317}, the interacting boson-fermion model \cite{ALONSO1995NPA100}, etc. Microscopic calculations for odd-mass nuclei are generally more complex because, in addition to collective degrees of freedom, one has to explicitly consider single-particle degrees of freedom. To this end, nuclear energy density functionals (EDFs) generally present a good starting point since EDFs enable a complete and accurate description of ground-state properties, collective excitations, and decay properties over the entire chart of nuclides \cite{Bender2003RMP121,Vretenar2005Phys.Rep.101,Meng2006Prog.Part.Nucl.Phys.470,STONE2007PPNP587,Niksic2011Prog.Part.Nucl.Phys.519,Mengbook2016,Robledo2018JPG013001,Zhao2011Phys.Lett.B181,Zhao2011Phys.Rev.Lett.122501,Niu2013PLB172,Niu2013PRC051303,Quan2017Phys.Rev.C54321,Lu2015Phys.Rev.C27304,ZhaoLi2018}. Recently several EDF-based approaches, e.g., the generator coordinate method (GCM) built on blocked one-quasiparticle states \cite{Bally2014PRL162501,Borrajo2018PRC044317,BORRAJO2017PLB328}, the interacting boson-fermion model (IBFM) \cite{Nomura2016Phys.Rev.C54305}, and the core-quasiparticle coupling (CQC) model \cite{Quan2017Phys.Rev.C54309} have been implemented for spectroscopic studies of odd-A quadrupole-deformed nuclei. Microscopic  models for odd-A nuclei with octupole deformation are much less developed. Only very recently the IBFM was extended to include both quadrupole and octupole boson degrees of freedom and applied to a study of low-lying positive- and negative-parity spectra of odd-A $^{143-147}$Ba \cite{Nomura2018Phys.Rev.C24317}.

In this work we extend the microscopic core-quasiparticle coupling model of Ref.\cite{Quan2017Phys.Rev.C54309} to include both quadrupole and octupole deformations in reflection-asymmetric odd-mass nuclear systems, based on the framework of covariant EDFs \cite{Vretenar2005Phys.Rep.101,Mengbook2016}. The major development is the inclusion of octupole correlations in both the even-even core Hamiltonian \cite{Li2013Phys.Lett.B866, Li2016J.Phys.G24005, Xia2017Phys.Rev.C54303, Xu2017Chin.Phys.C124107}, with parameters determined by the axially reflection-asymmetric relativistic Hartree-Bogoliubov model calculations, and the core-quasiparticle coupling terms. Also, the inclusion of both neighboring even-even cores in the configuration space enables the model to take into account shape polarization effects, that is, differences in shapes and related observables which are critical for transitional nuclei \cite{Quan2018Phys.Rev.C31301}. The present study is focused on the low-lying positive- and negative-parity spectra of odd-A $^{223, 225, 227}$Ra, located in the vicinity of the even-even nucleus $^{224}$Ra with static octupole deformation. Furthermore, $^{225}$Ra is a favorable case to search for a permanent electric dipole moment.

In Sec.~\ref{secII} the model and self-consistent microscopic calculation of the parameters of the Hamiltonian are described. Illustrative results for spectroscopic properties, such as low-lying excitation spectra, electric transitions, and wave functions are discussed in Sec.~\ref {secIII}. Section \ref{secIV} contains a summary of the principal results.

\section{\label{secII} Theoretical framework}

\subsection{Core-quasiparticle coupling model with quadrupole and octupole interactions}

In the core-quasiparticle coupling scheme, the configuration space of the odd-A nucleus is composed of both a particle coupled to the lighter even-even neighbor $A-1$ and a hole coupled to the heavier even-even neighbor $A+1$. For an odd-A nucleus in which low-energy excitation spectra are determined by octupole correlations, both the ground-state bands and the lowest negative-parity bands of the even-even cores are included in the configuration space. The ansatz for the wave function  of the odd-mass system can therefore be written in the following form:
\begin{equation}
\label{eq:wf}
|\alpha JM_J\pi\rangle^A = \sum_{\mu\Omega} \Big\{ U_{\mu\Omega}\big[a^\dag_{\mu}|\Omega\rangle\big]^{A-1}_{JM_J\pi} +V_{\mu\Omega}\big[a_{\mu}|\Omega\rangle\big]^{A+1}_{JM_J\pi}\Big\},
\end{equation}
where the total angular momentum is $J$, its projection is $M_J$, $\pi$ denotes the parity, and $\alpha$ denotes all additional quantum numbers that characterize the state. $\mu$ denotes the angular momentum $j$, its projection $m_j$, parity $\pi_j$, and additional quantum numbers $n_\mu$ of single-particle states, while $\Omega$ is the generic notation for the angular momentum $R$, its projection $M_R$, parity $\pi_R$, and additional quantum numbers $n_\Omega$ of collective states of the even-even core.
The coefficients $U_{\mu\Omega}$ and $V_{\mu\Omega}$ represent the probability amplitudes for the particle-like and hole-like states, respectively, that are formed by coupling the spherical single-particle state $|\mu\rangle^{A-1}$ to the collective state $|\Omega\rangle^{A-1}$ of the core $A-1$, and the spherical single-hole state $|\mu\rangle^{A+1}$ to the corresponding collective state $|\Omega\rangle^{A+1}$ of the core $A+1$.

The Hamiltonian of the CQC model reads
\begin{align}
H &=H_{\rm qp}+H_{\rm c}  \nonumber\\
   &=\left(
\begin{array}{cc}
(\varepsilon^{A-1}-\varepsilon_f)+\Gamma^{A-1} & \Delta^{A+1} \\
\Delta^{\dag A-1} & -(\varepsilon^{A+1}-\varepsilon_f)-\Gamma^{A+1}
\end{array}\right) \nonumber\\
& \ \ \ +\left(
\begin{array}{cc}
E^{A-1} & 0 \\
0 & E^{A+1}
\end{array}\right).
\label{eq:Ham}
\end{align}
The matrices $(\varepsilon^{A\pm1}-\varepsilon_f)$ and $(E^{A\pm1})$ are diagonal with respect to the basis states in the decomposition (\ref{eq:wf}):
\begin{align}
\label{eq:spe}
(\varepsilon^{A\pm1}-\varepsilon_f) &= (\varepsilon_{\mu}^{A\pm1}-\varepsilon_f)\delta_{\mu\mu^\prime}\delta_{\Omega\Omega^\prime}, \\
(E^{A\pm1}) &= E_{\Omega}^{A\pm1}\delta_{\mu\mu^\prime}\delta_{\Omega\Omega^\prime}
\end{align}
with the spherical single-particle energies $\varepsilon_{\mu}^{A\pm1}$, Fermi surface $\varepsilon_f$, and collective excitation energies $E_{\Omega}^{A\pm1}$. $\Gamma$ and $\Delta$ denote the mean field and pairing field that correspond to the long-range particle-hole and short-range particle-particle interactions, respectively.  The quadrupole-quadrupole and octupole-octupole interactions determine the fields $\Gamma$, and a monopole pairing force is used for $\Delta$:
\begin{align}
\label{eq:Gamm}
\left(\Gamma^{A\pm1}\right) &=-\chi_2 (-1)^{\small{j+R+J}}\small{\left\{\begin{array}{ccc} j & 2 & j^\prime \\ R^\prime & J & R \end{array}\right\} }
                              \left[\langle\mu\|\hat Q_2\|\mu^\prime\rangle\langle\Omega\|\hat Q_2\|\Omega^\prime\rangle\right]^{A\pm1} \nonumber\\
                              & \ \ \ -\chi_3 (-1)^{\small{j+R+J}}\small{\left\{\begin{array}{ccc} j & 3 & j^\prime \\ R^\prime & J & R \end{array}\right\}}
                              \left[\langle\mu\|\hat Q_3\|\mu^\prime\rangle\langle\Omega\|\hat Q_3\|\Omega^\prime\rangle\right]^{A\pm1}, \\
\left(\Delta^{A+1}\right) &=\left(\Delta^{A-1}\right) =\langle\Omega, A-1|\hat\Delta|\Omega^\prime, A+1\rangle \delta_{\mu\mu^\prime} \nonumber \\
                                  & \ \ \ \ \ \ \ \ \ \ \ \ \ \   \approx \frac{1}{2}(\Delta^{A-1}_{\Omega}+\Delta^{A+1}_{\Omega})
                                                                        \delta_{\mu\mu^\prime}\delta_{\Omega\Omega^\prime}\equiv\left(\Delta\right),
\end{align}
where $\langle\mu\|\hat Q_\lambda\|\mu^\prime\rangle^{A\pm1}$ and $\langle\Omega\|\hat Q_\lambda\|\Omega^\prime\rangle^{A\pm1}$ are the reduced quadrupole ($\lambda=2$) and octupole ($\lambda=3$) matrix elements of the spherical single-nucleon and core states, respectively. $\chi_2$ and $\chi_3$ are the corresponding coupling strengths of the quadrupole and octupole  fields. $\Delta_{\Omega}^{A\pm1}$ denotes the average pairing fields in the collective states $|\Omega\rangle^{A\pm1}$.

The Fermi surface $\varepsilon_f$ and coupling strengths $\chi_2$ and $\chi_3$ are free parameters that are adjusted to reproduce the experimental ground-state spin and parity and/or the excitation energies of few lowest levels. Finally, the excitation energies $E_{\alpha J\pi}$ and eigenvectors $U_{\mu\Omega}$, $V_{\mu\Omega}$ are obtained as solutions of the eigenequation
\begin{widetext}
\begin{equation}
\label{eq:hamo}
\left(
\begin{array}{cc}
(\varepsilon^{A-1}-\varepsilon_f)+\Gamma^{A-1}+E^{A-1} & \Delta \\
\Delta & -(\varepsilon^{A+1}-\varepsilon_f)-\Gamma^{A+1}+E^{A+1}
\end{array}\right)
\left(
\begin{array}{c}
U \\ V
\end{array}\right)
=E_{\alpha J\pi}
\left(
\begin{array}{c}
U \\ V
\end{array}\right)
\end{equation}
\end{widetext}
following the method introduced in Ref.~\cite{Klein2004PRC034338}.

In its current version the CQC Hamiltonian Eq. (\ref{eq:Ham}) is formulated in the laboratory frame of reference. The deformation of single-nucleon orbitals is not static, rather it is induced dynamically by the quadrupole and octupole coupling terms in Eq. (\ref{eq:Gamm}). In the Appendix of Ref.\cite{Frauendorf2014PRC014322} an example for transforming the quadrupole coupling term to the principal-axis system of the core (intrinsic frame) is discussed (cf. Eq. (A2) in Ref. \cite{Frauendorf2014PRC014322}). By combining the resulting deformed potential with the spherical quasiparticle Hamiltonian, one obtains the familiar form of the quasiparticle rotor model. Similarly, in the model used in the present work the quadrupole-quadrupole and octupole-octupole interactions between the core nucleus and the odd-nucleon in Eq. (\ref{eq:Gamm}) will play the same role and induce the right deformation for the odd-nucleon orbitals.

The electromagnetic multipole operator consists of single-nucleon and collective parts:
\begin{equation}
\hat{M}_{\lambda\nu}=\hat{M}^{\text{s.p.}}_{\lambda\nu}+\hat{M}^{\text{c}}_{\lambda\nu}.
\end{equation}
The corresponding reduced transition matrix element between states $|\alpha_iJ_i\pi_i\rangle$
and $|\alpha_fJ_f\pi_f\rangle$ reads
\begin{widetext}
\begin{align}
 \langle \alpha_fJ_f\pi_f\|\hat{M}_{\lambda}\|\alpha_iJ_i\pi_i\rangle
 &=\sqrt{(2J_f+1)(2J_i+1)}\sum_{\Omega,\mu_f\mu_i}(-1)^{j_f+J_i+R+\lambda}
 \left\{\begin{array}{ccc}
         J_f & \lambda & J_i \\
         j_i & R       & j_f
 \end{array}\right\}
 \Big[{M}^{\text{s.p.}}_U+{M}_V^{\text{s.p.}}\Big]\nonumber \\
&+\sqrt{(2J_f+1)(2J_i+1)}\sum_{\mu,\Omega_f\Omega_i}(-1)^{R_f+J_i+j+\lambda}
 \left\{\begin{array}{ccc}
         J_f & \lambda & J_i \\
         R_i & j       & R_f
 \end{array}\right\}
\Big[{M}^{\text{c}}_U+{M}_V^{\text{c}}\Big]
\end{align}
\end{widetext}
with
\begin{align}
\label{eq:MUsp}
M_U^{\text{s.p.}}&=U^f_{\mu_f\Omega}U^i_{\mu_i\Omega}
                                \langle \mu_f\|\hat{M}^{\text{s.p.}}_{\lambda}\|\mu_i\rangle^{A-1}, \\
M_V^{\text{s.p.}}&=V^f_{\mu_f\Omega}V^i_{\mu_i\Omega}
                                \langle \mu_f\|\hat{M}^{\text{s.p.}}_{\lambda}\|\mu_i\rangle^{A+1}, \\
M_U^{\text{c}}   &=U^f_{\mu\Omega_f}U^i_{\mu\Omega_i}
                               \langle \Omega_f\|\hat{M}^{\text{c}}_{\lambda}\|\Omega_i\rangle^{A-1}, \\
\label{eq:MVc}
M_V^{\text{c}}   &=V^f_{\mu\Omega_f}V^i_{\mu\Omega_i}
                               \langle \Omega_f\|\hat{M}^{\text{c}}_{\lambda}\|\Omega_i\rangle^{A+1}.
\end{align}
The explicit expressions for the reduced matrix elements of the electric dipole, quadrupole, and octupole operators are given  in Sec. \ref{SecIIB}.

\subsection{\label{SecIIB} Microscopic inputs based on covariant EDFs}

The full dynamics of the CQC Hamiltonian Eq.(\ref{eq:Ham}) is determined by the energies $\varepsilon_{\mu}^{A\pm1}$ and $E^{A\pm1}_{\Omega}$, quadrupole matrix elements $\langle\mu\|\hat Q_2\|\mu^\prime\rangle^{A\pm1}$ and $\langle\Omega\|\hat Q_2\|\Omega^\prime \rangle^{A\pm1}$, octupole matrix elements   $\langle\mu\|\hat Q_3\|\mu^\prime\rangle^{A\pm1}$ and $\langle\Omega\|\hat Q_3\|\Omega^\prime \rangle^{A\pm1}$, and pairing gaps $\Delta_{\Omega}^{A\pm1}$. In the following the superscript $A\pm1$  will be omitted for convenience.  In this particular study we will calculate the input for the CQC model using an axially reflection-asymmetric implementation of the relativistic Hartree-Bogoliubov (RHB) model \cite{Marevic2018Phys.Rev.C24334}, combined with the quadrupole-octupole collective Hamiltonian \cite{Xia2017Phys.Rev.C54303}. The RHB model provides a unified description of particle-hole $(ph)$ and particle-particle $(pp)$ correlations on a mean-field level by combining two average potentials: the self-consistent mean field that encloses long range \textit{ph} correlations and a pairing field $\hat{\Delta}$ which sums up \textit{pp} correlations.  In the present analysis, the mean-field potential is determined by the relativistic density functional PC-PK1 \cite{Zhao2010Phys.Rev.C54319} in the $ph$ channel, and a separable pairing force~\cite{Tian2009Phys.Lett.B44} is used in the $pp$ channel. PC-PK1  denotes a parametrization for the covariant nuclear energy density functional with nonlinear point-coupling interactions, that was adjusted in a fit to ground-state properties of 60 selected spherical nuclei, including binding energies, charge radii, and empirical pairing gaps  \cite{Zhao2010Phys.Rev.C54319} .

In the first step of the construction of the CQC Hamiltonian (\ref{eq:Ham}), a constrained RHB calculation for spherical configurations of the even-mass cores is carried out to determine the single-nucleon states  $|\mu\rangle$ and corresponding  energies $\varepsilon_{\mu}$, and the quadrupole ($\lambda=2$) and octupole ($\lambda=3$) matrix elements
\begin{align}
\label{eq:spq}
\langle\mu\|\hat Q_\lambda\|\mu^\prime\rangle &=\langle n_\mu j\|r^\lambda Y_\lambda\|n_\mu^\prime j^\prime\rangle   \nonumber\\
      &= (-1)^{j+j^\prime+\lambda+1}\sqrt{\frac{(2\lambda+1)(2j^\prime+1)}{4\pi}} \nonumber\\
      & \ \ \ \times C^{j\frac{1}{2}}_{j^\prime\frac{1}{2} \lambda0}\langle n_\mu j|r^\lambda|n_\mu^\prime j^\prime\rangle,
\end{align}
where $C^{j\frac{1}{2}}_{j^\prime\frac{1}{2} \lambda0}$ are the corresponding Clebsch-Gordan coefficients.

In the next step a constrained, axially reflection-asymmetric RHB calculation for the entire energy surface as function of the quadrupole deformation $\beta_2$ and ocutpole deformation $\beta_3$ is performed to compute the microscopic input, {\it i.e.}, the moments of inertia $\mathcal{I}$, collective masses $B_{22}$, $B_{23}$, $B_{33}$, and the potential $V_{\textnormal{coll}}$ for the quadrupole-octupole collective Hamiltonian (QOCH) of the even-mass core \cite{Xia2017Phys.Rev.C54303}:
\begin{equation}
\begin{split}
{{\hat{H}}_{\text{coll}}}=&-\dfrac{\hbar^2}{2\sqrt{w\mathcal{I}}}
          \left[\dfrac{\partial}{\partial \beta_2}\sqrt{\dfrac{\mathcal{I}}{w}}B_{33}\dfrac{\partial}{\partial \beta_2}
         -\dfrac{\partial}{\partial \beta_2}\sqrt{\dfrac{\mathcal{I}}{w}}B_{23}\dfrac{\partial}{\partial \beta_3}\right. \\
       &\left.-\dfrac{\partial}{\partial \beta_3}\sqrt{\dfrac{\mathcal{I}}{w}}B_{23}\dfrac{\partial}{\partial \beta_2}
         +\dfrac{\partial}{\partial \beta_3}\sqrt{\dfrac{\mathcal{I}}{w}}B_{22}\dfrac{\partial}{\partial \beta_3}\right] \\
       &+\dfrac{\hat{J}_\perp^2}{2\mathcal{I}}+{{V}_{\text{coll}}}( {{\beta }_{2}}, {{\beta }_{3}} ).
\end{split}
   \label{eq:CH}
\end{equation}
$\hat{J}_\perp$ denotes the component of angular momentum perpendicular to the symmetric axis in the body-fixed frame of the core. The Coriolis interaction is contained in the rotational term of QOCH because of $J_\perp=J-j$, where $J$ and $j$ are the total angular momentum of the odd-A nucleus and angular momentum of the single nucleon, respectively. In addition to the rotational term, we also include the vibrational term and the collective potential in Eq. (\ref{eq:CH}), and this enables a description of more complex excitation modes, e.g., phonon excitations and shape coexistence. The additional quantity that appears in the vibrational kinetic energy, $w=B_{22}B_{33}-B_{23}^2$, determines the volume element in the collective space,
\begin{equation}
\int d\tau_\text{coll}=\int\sqrt{w\cal{I}}d\beta_2 d\beta_3 d\Theta,
\end{equation}
where $\Theta$ are the Euler angles.

The eigenvalue problem of the collective Hamiltonian \eqref{eq:CH} is solved using an expansion of eigenfunctions in terms of a complete set of basis functions that depend on the collective coordinates $\beta_2$, $\beta_3$, and $\Theta$ \cite{Xia2017Phys.Rev.C54303}. The collective wave functions are thus obtained as
\begin{equation}
\Psi _{\Omega}\left( {{\beta }_{2}},{{\beta }_{3}},\Theta \right)=\psi _{n_\Omega R\pi_R  }\left( {{\beta }_{2}},{{\beta }_{3}} \right)\left| RM_R0 \right\rangle.
  \label{eq:wave}
\end{equation}
The corresponding reduced matrix elements and average pairing gaps are calculated using the expressions
\begin{align}
\label{eq:cq}
\langle\Omega\|\hat Q_\lambda\|\Omega^\prime\rangle &=\sqrt{2R+1}C_{R^\prime 0 \lambda0}^{R0}  \nonumber\\
 & \ \ \ \times\int d\beta_2 d\beta_3\sqrt{w\cal{I}}\psi _{n_\Omega^\prime R^\prime\pi_R^\prime}q_\lambda(\beta_2, \beta_3)\psi_{n_\Omega R\pi_R}^{* } \\
\Delta_\Omega &=\int d\beta_2 d\beta_3\sqrt{w\cal{I}}|\psi_{n_\Omega R\pi_R}|^2\Delta(\beta_2, \beta_3),
\end{align}
where $q_{\lambda}(\beta_2, \beta_3)$ and $\Delta(\beta_2, \beta_3)$ are the mass quadrupole ($\lambda=2$) and octupole ($\lambda=3$) moments
and pairing gaps, respectively, computed from the RHB intrinsic state at each point on the deformation surface $(\beta_2, \beta_3)$.

In the case of electric dipole transitions the reduced matrix elements for the single-nucleon states are calculated from:
\begin{align}
\label{eq:spd1}
\langle\mu\|\hat D_1\|\mu^\prime\rangle &=
    e_\tau (-1)^{j+j^\prime}\sqrt{\frac{3(2j^\prime+1)}{4\pi}}C^{j\frac{1}{2}}_{j^\prime\frac{1}{2} 10}\langle n_\mu j|r|n_\mu^\prime j^\prime\rangle
\end{align}
with $e_\tau=\frac{N}{A}e$ for the odd proton and $e_\tau=-\frac{Z}{A}e$ for the odd neutron.
For the even-even core:
\begin{align}
\label{eq:D1c}
\langle\Omega\|\hat D_1\|\Omega^\prime\rangle &=\sqrt{2R+1}C_{R^\prime 0 10}^{R0} \nonumber \\
 &\ \ \  \times\int d\beta_2 d\beta_3\sqrt{w\cal{I}}\psi _{n_\Omega^\prime R^\prime\pi_R^\prime}D_1(\beta_2, \beta_3)\psi_{n_\Omega R\pi_R}^{* },
\end{align}
where $D_1(\beta_2, \beta_3)$ is the electric dipole moment calculated from the RHB intrinsic state using the dipole operator
\begin{equation}
\hat D_1=\sqrt{\frac{3}{4\pi}}e\left(\frac{N}{A}z_p-\frac{Z}{A}z_n\right).
\end{equation}


\section{\label{secIII}Results and Discussion}

\begin{figure}[htbp]
\includegraphics[scale=0.62]{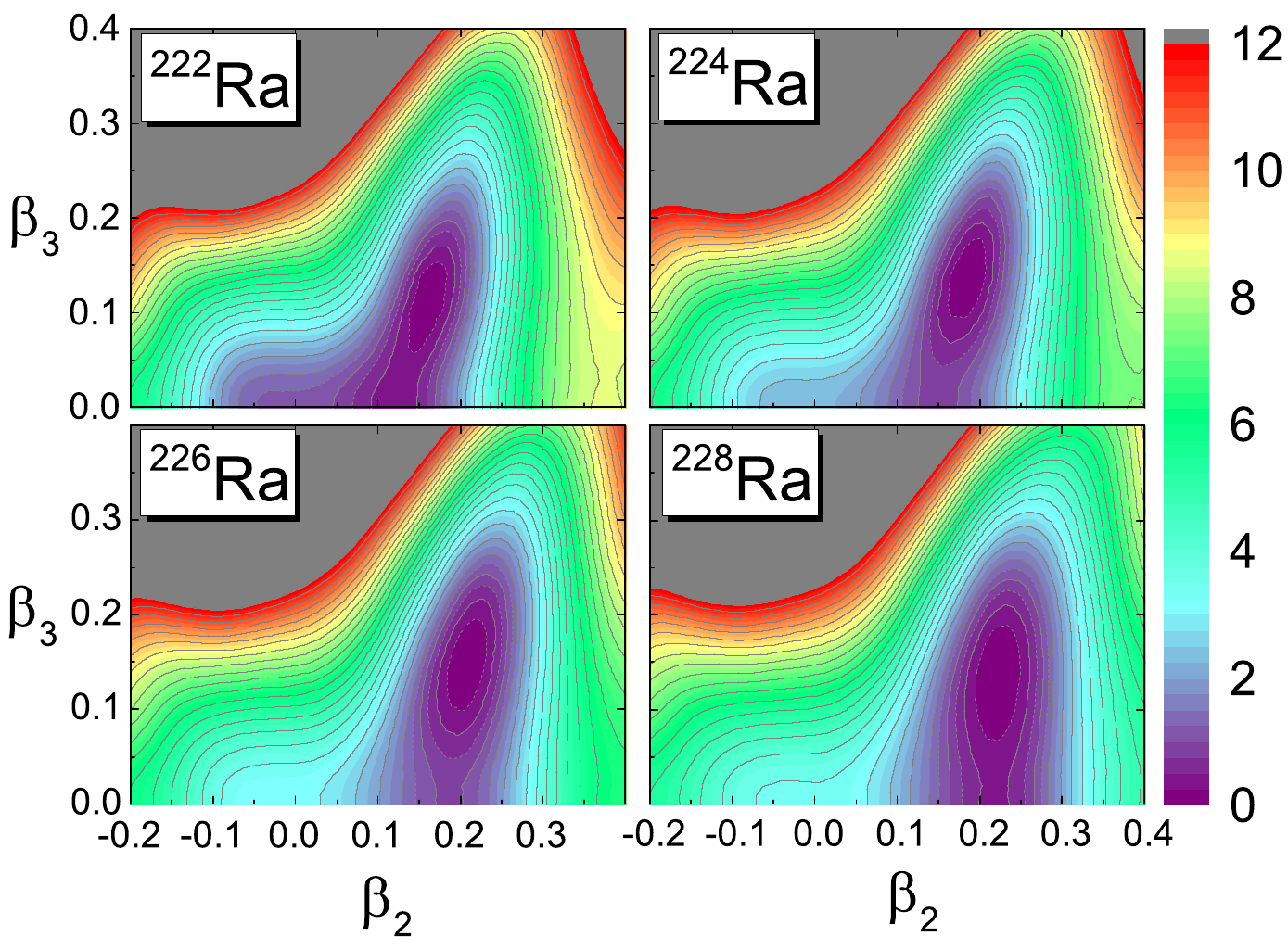}
\caption{\label{fig:core-pes}Self-consistent RHB deformation energy surfaces of the core nuclei $^{222-228}$Ra in the $\beta_2$-$\beta_3$ plane. For each nucleus the energies are normalized with respect to the binding energy of the global minimum. The contours join points on the surface with the same energy (in MeV). Contour lines are separated by 0.5 MeV (between neighboring solid lines) and 0.25 MeV (between neighboring dashed and solid lines), respectively.}
\end{figure}

\begin{figure*}[htbp]
\includegraphics[scale=1.3]{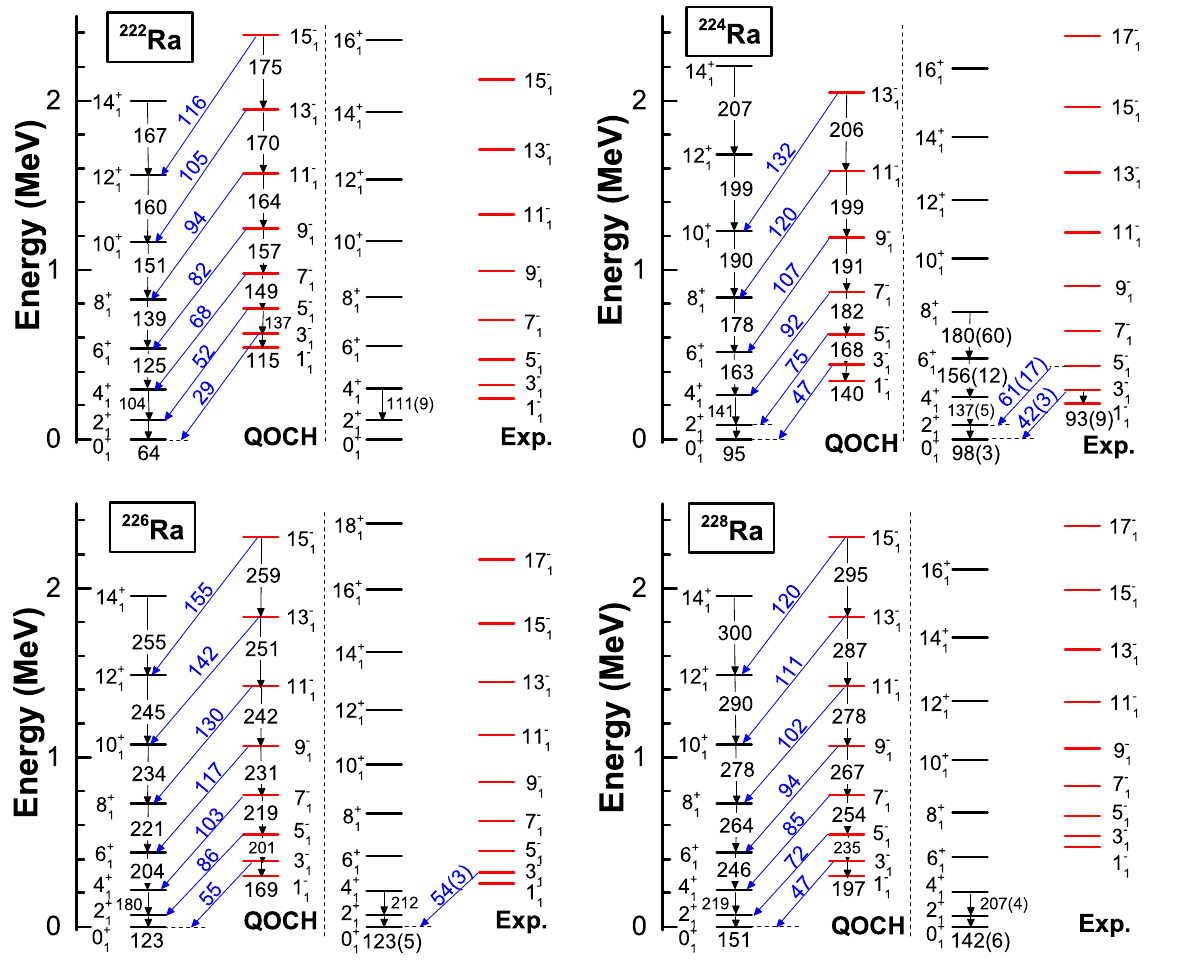}
\caption{The excitation spectrum, intraband $B(E2)$ values (in W.u.), and interband $B(E3)$ values (in W.u.) of the core nuclei $^{222-228}$Ra calculated with the QOCH based on the PC-PK1 relativistic density functional and compared to experimental results \cite{NNDC,Gaffney2013Nature199}.}
\label{fig:even-bgh}
\end{figure*}

As an illustrative application of the microscopic CQC model, we consider the case of a single neutron coupled to an axially symmetric octupole-deformed even core: the spectroscopy of $^{223, 225, 227}$Ra, for which extensive data are available. Furthermore, we note that $^{225}$Ra is the nucleus of choice in the search for a permanent electric dipole moment. The corresponding even-core nuclei $^{222-228}$Ra present good examples of axially symmetric octupole-deformed systems, as shown by the deformation energy surfaces in Fig. \ref{fig:core-pes}. All four core nuclei $^{222-228}$Ra exhibit a global minimum with $\beta_3>0.1$ and octupole deformation energy  $\Delta E_{\rm oct}>0.4$ MeV.  As the neutron number increases, both the octupole deformation and $\Delta E_{\rm oct}$ increase from $^{222}$Ra to $^{226}$Ra, and then start to decrease from $^{228}$Ra. The isotope $^{226}$Ra exhibits the deepest octupole minimum with $\Delta E_{\rm oct}\approx0.94$ MeV. Similar deformation energy surfaces of Ra isotopes have also been obtained in the RHB calculation with the DD-PC1 functional \footnote{DD-PC1 denotes a parametrization for the covariant nuclear energy density functional with density-dependent point-coupling interactions. It was adjusted in a fit to the experimental masses of a set of 64 deformed nuclei in the mass regions $A\approx150-180$ and $A\approx230-250$ \cite{Niksic2008Phys.Rev.C34318}.}, but in that case the most pronounced octupole minimum has been obtained for $^{224}$Ra \cite{Nomura2014Phys.Rev.C24312}.

The solution of the eigenvalue equations for the corresponding quadrupole-octupole collective Hamiltonian yields the collective excitation spectra of the even-core nuclei. Figure \ref{fig:even-bgh} displays the resulting excitation energies, intraband $B(E2)$ values, and interband $B(E3)$ values for the ground-state bands and lowest-lying negative-parity bands.  Except for the calculated negative-parity band heads in $^{222}$Ra and $^{228}$Ra being somewhat higher and lower than the corresponding experimental counterparts, respectively, the overall structure agrees very well with the available data. It should be emphasized that the electric transition rates are in excellent agreement with data. This is relevant for the calculation of odd-A nuclei using the CQC model, because it includes the reduced quadrupole and octupole matrix elements as input.

The next step is the construction of the CQC Hamiltonian. The fermion space consists of spherical single-neutron states with $E^{sph}_{f}-2 \hbar\omega \leq \varepsilon_\mu\leq2$ MeV, where $E^{sph}_{f}$ is the Fermi surface of the corresponding spherical configuration and $\hbar\omega=41$$A^{-\frac{1}{3}}$ MeV. Both positive- and negative-parity single-neutron states in the canonical basis are included. For the collective states of the core nuclei, we include the following states: ${0^{+}_{1}}$, ${2^{+}_{1}}$,$\cdots$, ${18^{+}_{1}}$ in the positive-parity ground-state band and ${1^{-}_{1}}$, ${3^{-}_{1}}$,$\cdots$, ${17^{-}_{1}}$ in the lowest negative-parity band.

\begin{figure}[htbp]
\includegraphics[scale=0.3]{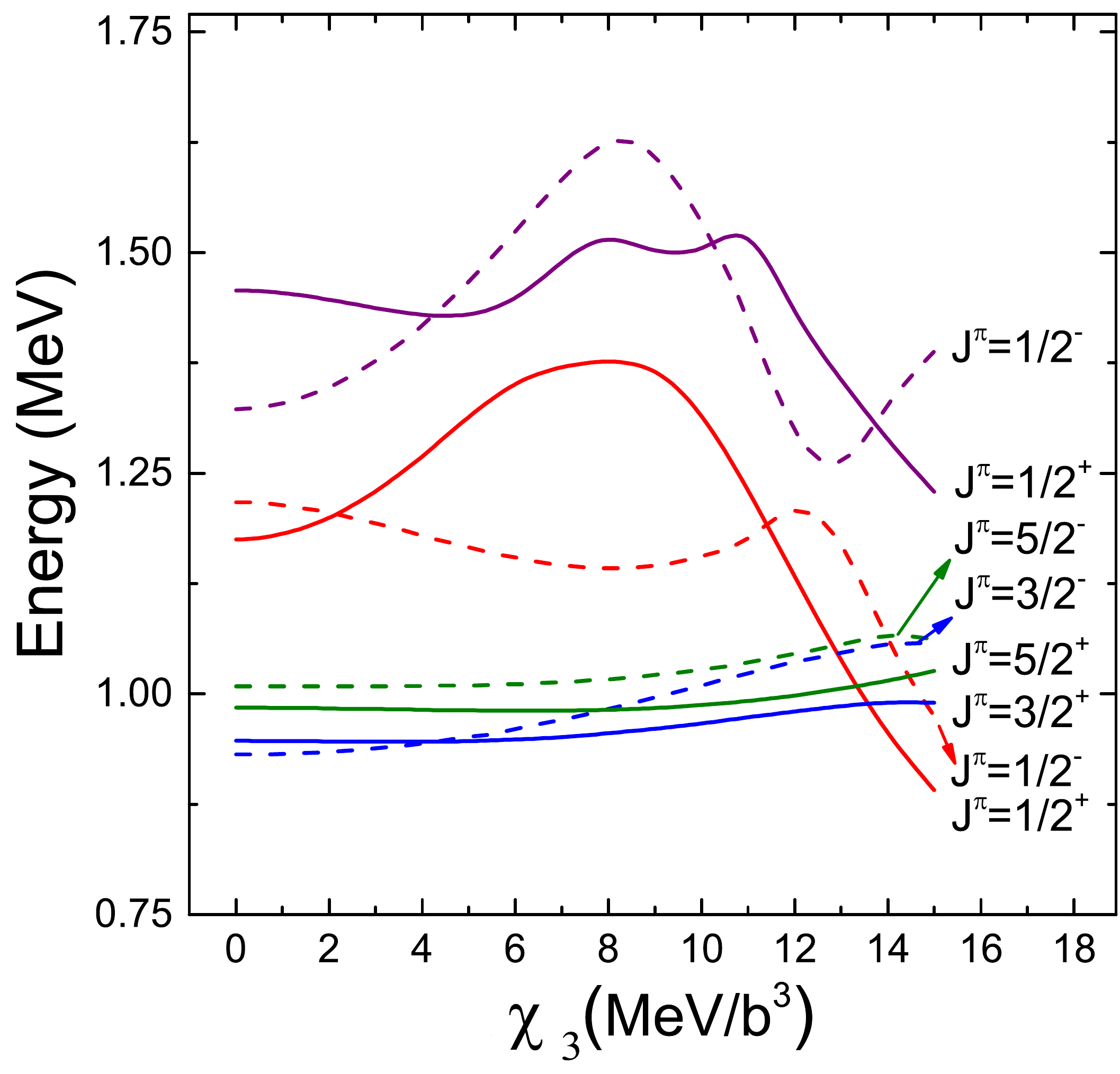}
\caption{
Quasiparticle energies [eigenenergies of $H_{\rm qp}$ in Eq. (\ref{eq:Ham})] relevant for the bandheads of the low-energy spectra of $^{223}$Ra: $J^\pi=1/2^\pm, 3/2^\pm$, and $5/2^\pm$, as functions of the octupole coupling strength  $\chi_3$ ($\chi_2=5.5$ MeV$/b^2$).
}
\label{fig:Echi3}
\end{figure}

Before discussing the full spectrum, in Fig. \ref{fig:Echi3} we show the evolution of quasiparticle energies [eigenenergies of $H_{\rm qp}$ in Eq. (\ref{eq:Ham})] relevant for the bandheads of the low-energy spectra of $^{223}$Ra: $J^\pi=1/2^\pm, 3/2^\pm$, and $5/2^\pm$, as functions of the octupole coupling strength $\chi_3$. One clearly observes the parity doublet structure of the bandheads. In particular, we note the pronounced dependence of the $J^\pi=1/2^\pm$ states on the strength of the octupole interaction.

\begin{figure*}[htbp]
\includegraphics[scale=1]{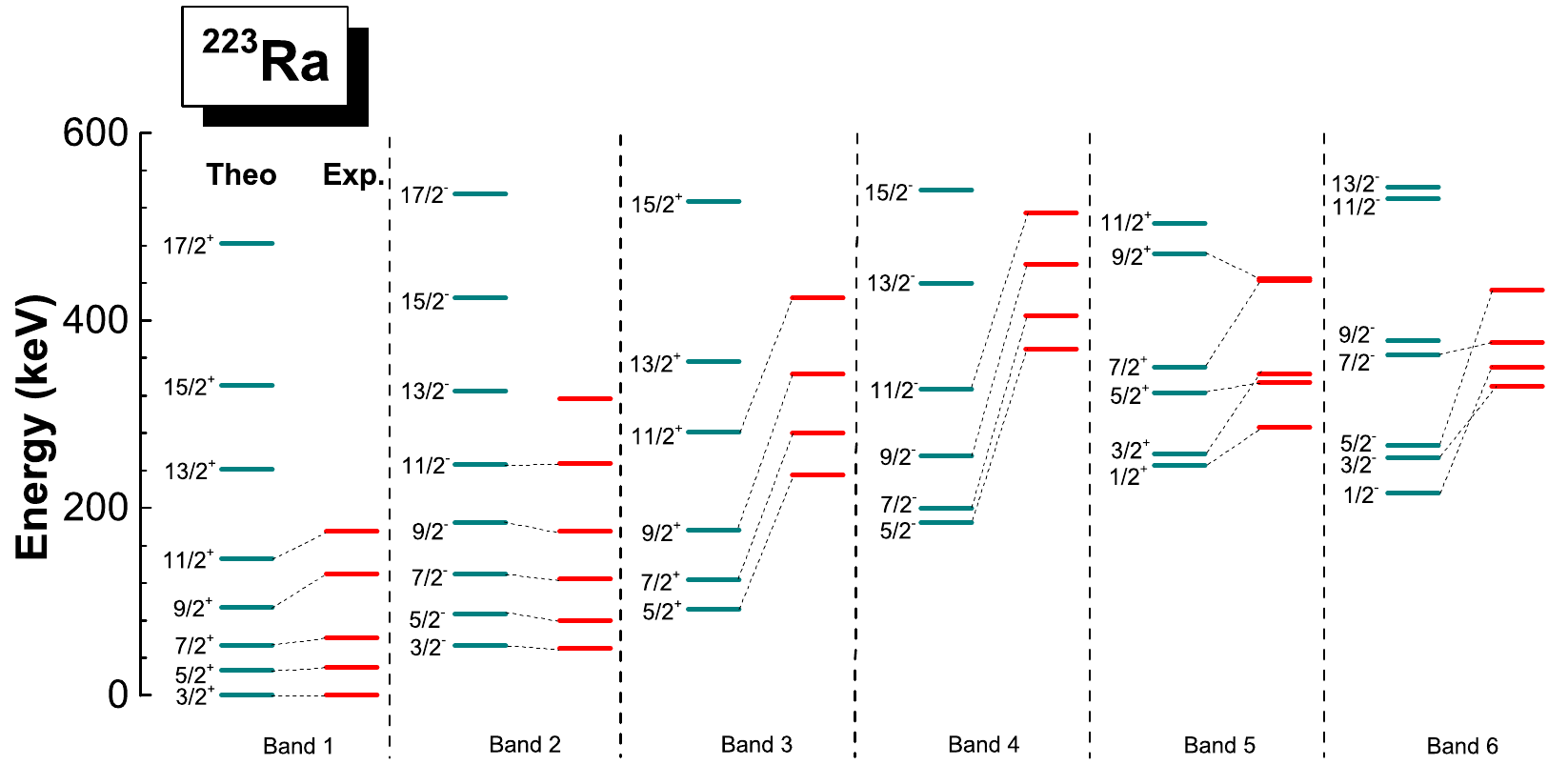}
\caption{The low-lying excitation spectrum of the odd-mass nucleus $^{223}$Ra calculated with the CQC model, plotted in comparison with available experimental data \cite{NNDC}.  The following values of the Fermi surface and coupling strengths are adjusted to low-energy data: $\varepsilon_f = -6.36$ MeV, $ \chi_2 = 5.5$ MeV/$b^2$, $ \chi_3 = 4.0$ MeV/$b^3$.}
\label{fig:bgh-223Ra}
\end{figure*}

On this and the following pages we group, by isotope, the figures with the excitation spectra and the tables with the dominant configurations. Figure \ref{fig:bgh-223Ra} displays the low-lying excitation spectrum of the odd-mass nucleus $^{223}$Ra calculated with the CQC model, in comparison with available experimental data \cite{NNDC}.  The levels are grouped into different bands according to the dominant decay pattern. Here six low-lying bands with bandheads $J^\pi=3/2^\pm$, $J^\pi=5/2^\pm$, and $J^\pi=1/2^\pm$ are shown, and they are displayed as three parity doublets. The ground-state band and its partner are well reproduced by the CQC model, especially the very small energy difference $\approx50$ keV between the two bandheads. In Table \ref{tab:wfRa223} we also list the probabilities of dominant configurations in the wave functions of selected states in the ground-state band and band 2 of $^{223}$Ra. The parity doublet is dominated by the strongly mixed configurations of $3d_{5/2}$ and $2g_{9/2}$ for the low-spin states, and gradually the $1i_{11/2}$ configuration becomes dominant for high-spin states. Note that both the ground-state band and band 2 are predominantly built by coupling the single neutron to the corresponding positive-parity and negative-parity bands in the core nuclei, respectively. There is no mixing between opposite-parity bands of the cores in the yrast states, and this points to weaker octupole correlations compared to $^{225}$Ra and $^{227}$Ra below. For the second parity doublet, i.e., bands 3 and 4, in Fig. \ref{fig:bgh-223Ra} the overall structure is reproduced but the theoretical bandheads are $\approx130$ keV lower than the corresponding experimental levels. We can also obtain the $J^\pi=1/2^\pm$ parity doublet and compare it with the experimental spectrum. However, the theoretical results exhibit opposite staggering, possibly due to a weaker Coriolis coupling, and this could be solved by adding a magnetic dipole particle-core interaction term \cite{Protopapas1997PRC1810}.

\begin{figure*}[htbp]
\includegraphics[scale=1]{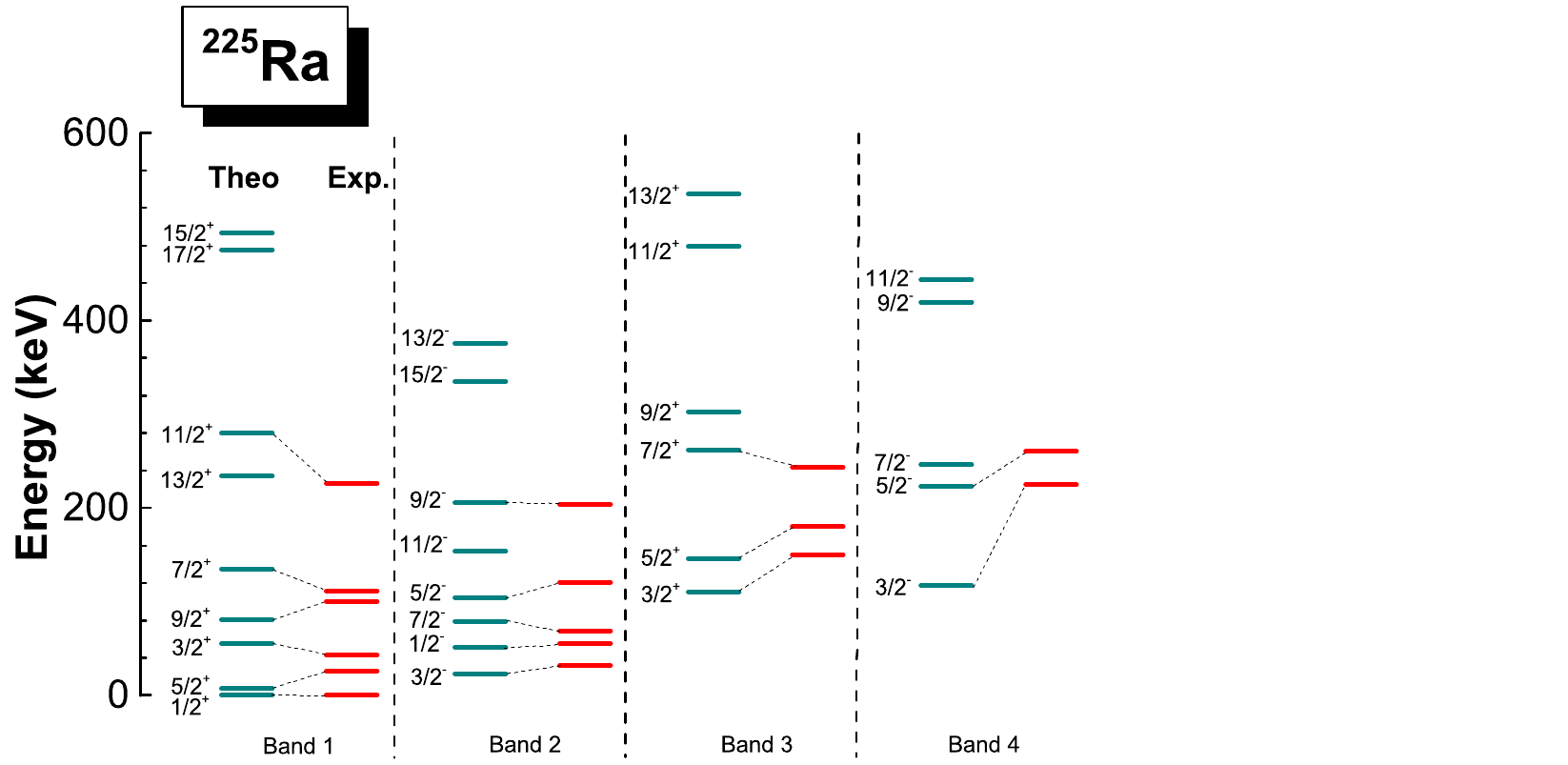}
\caption{ The same as in Fig. \ref{fig:bgh-223Ra} but for $^{225}$Ra. The Fermi surface and coupling strengths read $\varepsilon_f = -4.56$ MeV, $ \chi_2 = 6.9$ MeV/$b^2$, $ \chi_3 = 22.8$ MeV/$b^3$.}
\label{fig:bgh-225Ra}
\end{figure*}

In Fig. \ref{fig:bgh-225Ra} and Table \ref{tab:wfRa225} we display the excitation spectrum and the dominant configurations
in the wave functions of states in $^{225}$Ra, respectively.  In general, the low-lying excitations and parity doublet structure are reproduced by the CQC model, except the staggering behavior of the ground-state $K=1/2$ band. The ground state $J^\pi=1/2^+$ and its parity partner $J^\pi=1/2^-$ in band 2 are particularly interesting because they are associated with the enhancement of the Schiff moment \cite{ParkerRH2015,Auerbach1996PRL4316}. Here we obtain a good description for both the excitation energy and electric transition rates, as shown in Table \ref{tab:E1E2}. In Table \ref{tab:wfRa225} the pronounced  mixing between positive- and negative-parity single-particle states and collective states of the cores reflects a strong octupole coupling between the odd neutron and the even-even cores.

\begin{figure*}[htbp]
\includegraphics[scale=1]{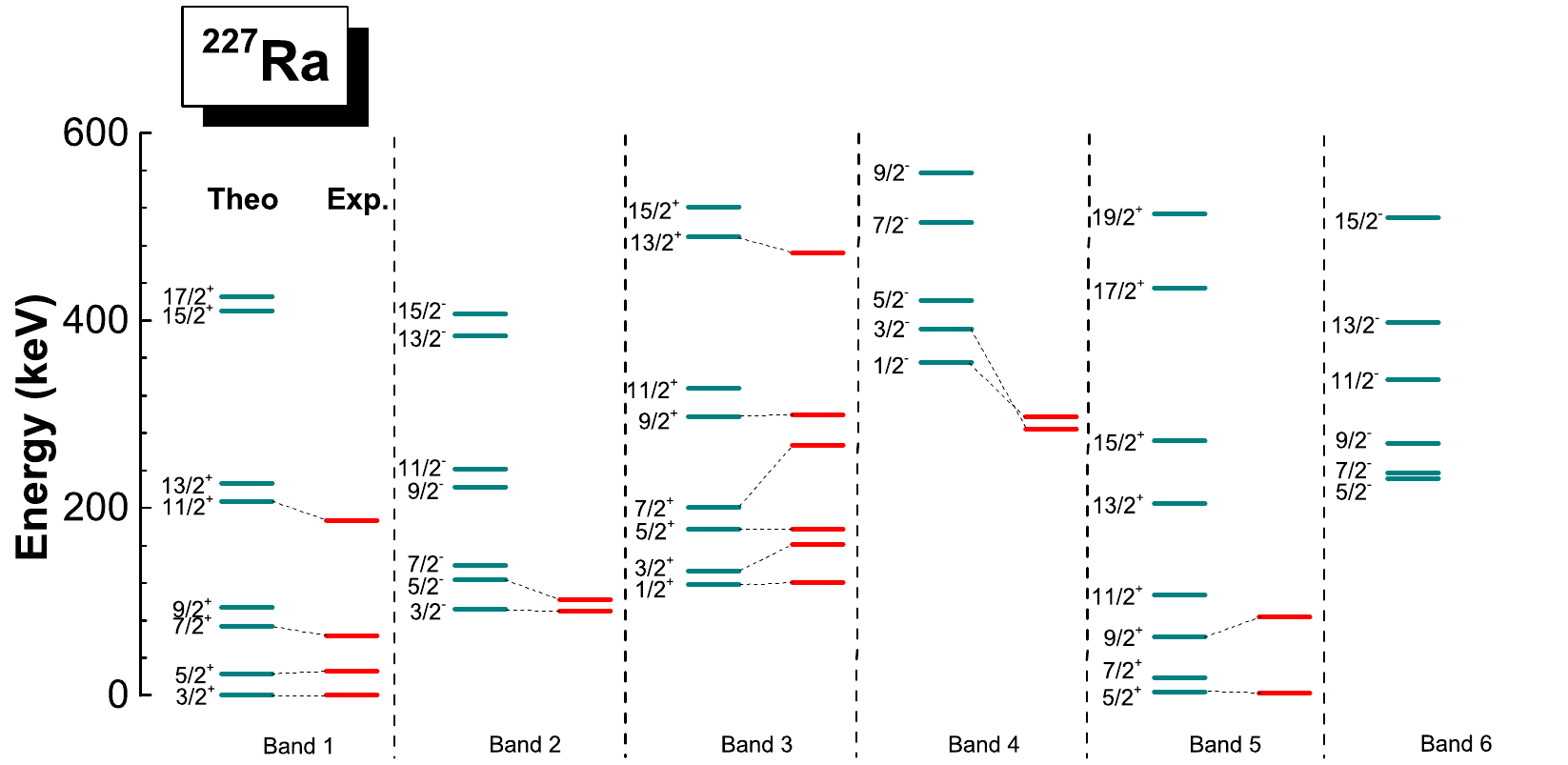}
\caption{The same as in Fig. \ref{fig:bgh-223Ra} but for $^{227}$Ra. The Fermi surface and coupling strengths read $\varepsilon_f = -6.28$ MeV, $ \chi_2 = 4.8$ MeV/$b^2$, $ \chi_3 = 12.5$ MeV/$b^3$.}
\label{fig:bgh-227Ra}
\end{figure*}

Figure \ref{fig:bgh-227Ra} and Table \ref{tab:wfRa227} contain the results for $^{227}$Ra. The structure of bands is similar to that found in $^{223, 225}$Ra. Bands 1, 2 and bands 3, 4 form two parity-doublet sequences. The low-lying band 5 is also reproduced but the partner band of opposite parity has not been determined in experiment. Band 6 is the corresponding theoretical prediction.  The ground-state band and its partner band in $^{227}$Ra are characterized by pronounced octupole correlations that are reflected in the mixing of positive- and negative-parity core states. One also notices that the lowest band structure of $^{227}$Ra is mainly built on the configurations with a neutron hole coupled to the heavier core $^{228}$Ra.

\begin{table*}[htbp]
\caption{\label{tab:wfRa223} The probabilities of the dominant configurations in the wave functions of selected states in the ground-state band and band 2 of $^{223}$Ra.}
\begin{ruledtabular}
\begin{tabular}{cccc|cccc}
$J^{\pi}$ & $j\otimes R$ & $A-1$ & $A+1$ & $J^{\pi}$ & $j\otimes R$ & $A-1$ & $A+1$ \\\colrule
$3/2^{+}$  & $3d_{5/2} \otimes 2_{1}^{+}$   &0.07&0.10      & $3/2^{-}$  & $3d_{5/2} \otimes 1_{1}^{-}$   &0.02&0.07\\
           & $2g_{9/2} \otimes 4_{1}^{+}$   &0.23&0.27      &            & $3d_{5/2} \otimes 3_{1}^{-}$   &0.03&0.07\\
           &                                &    &          &            & $2g_{9/2} \otimes 3_{1}^{-}$   &0.07&0.15\\
           &                                &    &          &            & $2g_{9/2} \otimes 5_{1}^{-}$   &0.10&0.22\\
           &                                &    &          &            & $1i_{11/2}\otimes 5_{1}^{-}$   &0.04&0.09\\
\hline
$5/2^{+}$  & $2g_{9/2} \otimes 2_{1}^{+}$   &0.12&0.11      & $5/2^{-}$  & $3d_{5/2} \otimes 3_{1}^{-}$   &0.02&0.09\\
           & $1i_{11/2}\otimes 4_{1}^{+}$   &0.08&0.16      &            & $2g_{9/2} \otimes 3_{1}^{-}$   &0.09&0.12\\
           & $2g_{9/2} \otimes 6_{1}^{+}$   &0.11&0.14      &            & $1i_{11/2}\otimes 3_{1}^{-}$   &0.02&0.09\\
           &                                &    &          &            & $2g_{9/2} \otimes 5_{1}^{-}$   &0.04&0.16\\
           &                                &    &          &            & $2g_{9/2} \otimes 7_{1}^{-}$   &0.02&0.06\\
\hline
$7/2^{+}$  & $1i_{11/2}\otimes 2_{1}^{+}$   &0.12&0.23      & $7/2^{-}$  & $3d_{5/2} \otimes 1_{1}^{-}$   &0.01&0.05\\
           & $2g_{9/2} \otimes 4_{1}^{+}$   &0.05&0.04      &            & $2g_{9/2} \otimes 1_{1}^{-}$   &0.06&0.08\\
           & $1i_{11/2}\otimes 4_{1}^{+}$   &0.08&0.02      &            & $1i_{11/2}\otimes 3_{1}^{-}$   &0.09&0.15\\
           & $2g_{9/2} \otimes 6_{1}^{+}$   &0.01&0.07      &            & $3d_{5/2} \otimes 5_{1}^{-}$   &0.03&0.05\\
           & $1i_{11/2}\otimes 6_{1}^{+}$   &0.01&0.11      &            & $1i_{11/2}\otimes 5_{1}^{-}$   &0.00&0.06\\
           &                                &    &          &            & $2g_{9/2} \otimes 7_{1}^{-}$   &0.04&0.17\\
\hline
$9/2^{+}$  & $1i_{11/2}\otimes 2_{1}^{+}$   &0.21&0.17      & $9/2^{-}$  & $1i_{11/2}\otimes 1_{1}^{-}$   &0.11&0.20\\
           & $1i_{11/2}\otimes 4_{1}^{+}$   &0.03&0.17      &            & $2g_{9/2} \otimes 5_{1}^{-}$   &0.04&0.06\\
           & $2g_{9/2} \otimes 8_{1}^{+}$   &0.02&0.06      &            & $1i_{11/2}\otimes 5_{1}^{-}$   &0.06&0.05\\
           & $1i_{11/2}\otimes 8_{1}^{+}$   &0.00&0.05      &            & $2g_{9/2} \otimes 7_{1}^{-}$   &0.00&0.07\\
           &                                &    &          &            & $1i_{11/2}\otimes 7_{1}^{-}$   &0.00&0.07\\
\hline
$11/2^{+}$ & $1i_{11/2}\otimes 0_{1}^{+}$   &0.19&0.20      & $11/2^{-}$ & $1i_{11/2}\otimes 1_{1}^{-}$   &0.09&0.06\\
           & $1i_{11/2}\otimes 2_{1}^{+}$   &0.01&0.09      &            & $1i_{11/2}\otimes 3_{1}^{-}$   &0.11&0.16\\
           & $1i_{11/2}\otimes 4_{1}^{+}$   &0.08&0.00      &            & $1i_{11/2}\otimes 5_{1}^{-}$   &0.00&0.10\\
           & $1i_{11/2}\otimes 6_{1}^{+}$   &0.04&0.08      &            & $2g_{9/2} \otimes 9_{1}^{-}$   &0.01&0.09\\
           & $1i_{11/2}\otimes 8_{1}^{+}$   &0.01&0.07      &            &                                &    &    \\
\hline
$13/2^{+}$ & $1i_{11/2}\otimes 2_{1}^{+}$   &0.19&0.11      & $13/2^{-}$ & $1i_{11/2}\otimes 1_{1}^{-}$   &0.16&0.22\\
           & $1i_{11/2}\otimes 4_{1}^{+}$   &0.08&0.17      &            & $1i_{11/2}\otimes 5_{1}^{-}$   &0.10&0.01\\
           & $1i_{11/2}\otimes 6_{1}^{+}$   &0.01&0.08      &            & $1i_{11/2}\otimes 7_{1}^{-}$   &0.01&0.08\\
           & $1i_{11/2}\otimes 10_{1}^{+}$  &0.01&0.06      &            &                                &    &    \\
\end{tabular}
\end{ruledtabular}
\end{table*}

\begin{table*}[htbp]
\caption{\label{tab:wfRa225} The same as in Table \ref{tab:wfRa223} but for $^{225}$Ra.}
\begin{ruledtabular}
\begin{tabular}{cccc|cccc}
$J^{\pi}$ & $j\otimes R$ & $A-1$ & $A+1$ & $J^{\pi}$ & $j\otimes R$ & $A-1$ & $A+1$ \\
\colrule
$1/2^{+}$  & $3d_{5/2} \otimes 2_{1}^{+}$   &0.18&0.12      & $1/2^{-}$  & $3d_{5/2} \otimes 3_{1}^{-}$   &0.18&0.11\\
           & $1i_{11/2}\otimes 6_{1}^{+}$   &0.11&0.06      &            & $2g_{7/2} \otimes 3_{1}^{-}$   &0.06&0.02\\
           & $1j_{15/2}\otimes 7_{1}^{-}$   &0.14&0.08      &            & $1i_{11/2}\otimes 5_{1}^{-}$   &0.10&0.06\\
           &                                &    &          &            & $1j_{15/2}\otimes 8_{1}^{+}$   &0.14&0.09\\
\hline
$3/2^{+}$  & $3d_{5/2} \otimes 2_{1}^{+}$   &0.09&0.08      & $3/2^{-}$  & $3d_{5/2} \otimes 1_{1}^{-}$   &0.14&0.10\\
           & $2g_{7/2} \otimes 4_{1}^{+}$   &0.07&0.05      &            & $1j_{15/2}\otimes 6_{1}^{+}$   &0.15&0.13\\
           & $1i_{11/2}\otimes 6_{1}^{+}$   &0.07&0.05      &            & $1i_{11/2}\otimes 7_{1}^{-}$   &0.09&0.05\\
           & $1j_{15/2}\otimes 7_{1}^{-}$   &0.14&0.15      &            &                                &    &    \\
\hline
$5/2^{+}$  & $3d_{5/2} \otimes 0_{1}^{+}$   &0.09&0.06      & $5/2^{-}$  & $3d_{5/2} \otimes 3_{1}^{-}$   &0.07&0.06\\
           & $3d_{5/2} \otimes 2_{1}^{+}$   &0.06&0.04      &            & $2g_{7/2} \otimes 5_{1}^{-}$   &0.06&0.06\\
           & $1j_{15/2}\otimes 5_{1}^{-}$   &0.14&0.11      &            & $1j_{15/2}\otimes 6_{1}^{+}$   &0.10&0.11\\
           & $1i_{11/2}\otimes 8_{1}^{+}$   &0.10&0.05      &            & $1i_{11/2}\otimes 7_{1}^{-}$   &0.05&0.04\\
           &                                &    &          &            & $1j_{15/2}\otimes 8_{1}^{+}$   &0.07&0.07\\
\hline
$7/2^{+}$  & $3d_{5/2} \otimes 4_{1}^{+}$   &0.06&0.05      & $7/2^{-}$  & $3d_{5/2} \otimes 1_{1}^{-}$   &0.11&0.07\\
           & $1j_{15/2}\otimes 5_{1}^{-}$   &0.08&0.07      &            & $1j_{15/2}\otimes 4_{1}^{+}$   &0.14&0.12\\
           & $2g_{7/2} \otimes 6_{1}^{+}$   &0.05&0.05      &            & $1i_{11/2}\otimes 9_{1}^{-}$   &0.09&0.04\\
           & $1j_{15/2}\otimes 7_{1}^{-}$   &0.05&0.07      &            &                                &    &    \\
\hline
$9/2^{+}$  & $3d_{5/2} \otimes 2_{1}^{+}$   &0.13&0.07      & $9/2^{-}$  & $1j_{15/2}\otimes 4_{1}^{+}$   &0.19&0.08\\
           & $1j_{15/2}\otimes 3_{1}^{-}$   &0.12&0.10      &            & $1i_{11/2}\otimes 7_{1}^{-}$   &0.14&0.02\\
           & $1i_{11/2}\otimes 10_{1}^{+}$  &0.09&0.04      &            &                                &    &    \\
\hline
$11/2^{+}$ & $3d_{5/2} \otimes 4_{1}^{+}$   &0.06&0.04      & $11/2^{-}$ & $1j_{15/2}\otimes 2_{1}^{+}$   &0.13&0.10\\
           & $1j_{15/2}\otimes 5_{1}^{-}$   &0.05&0.05      &            & $3d_{5/2} \otimes 3_{1}^{-}$   &0.13&0.06\\
           &                                &    &          &            & $1i_{11/2}\otimes 11_{1}^{-}$  &0.08&0.03\\
\hline
$13/2^{+}$ & $1j_{15/2}\otimes 1_{1}^{-}$   &0.09&0.07      & $13/2^{-}$ & $1j_{15/2}\otimes 2_{1}^{+}$   &0.11&0.03\\
           & $1j_{15/2}\otimes 3_{1}^{-}$   &0.05&0.05      &            & $1j_{15/2}\otimes 4_{1}^{+}$   &0.07&0.05\\
           & $3d_{5/2} \otimes 4_{1}^{+}$   &0.14&0.06      &            & $1i_{11/2}\otimes 9_{1}^{-}$   &0.12&0.01\\
           & $1i_{11/2}\otimes 12_{1}^{+}$  &0.09&0.03      &            &                                &    &    \\
\end{tabular}
\end{ruledtabular}
\end{table*}

\begin{table*}[htbp]
\caption{\label{tab:wfRa227} The same as in Table \ref{tab:wfRa223} but for $^{227}$Ra.}
\begin{ruledtabular}
\begin{tabular}{cccc|cccc}
$J^{\pi}$ & $j\otimes R$ & $A-1$ & $A+1$ & $J^{\pi}$ & $j\otimes R$ & $A-1$ & $A+1$ \\
\colrule
$3/2^{+}$  & $3p_{3/2} \otimes 1_{1}^{-}$   &0.02&0.11      & $3/2^{-}$  & $3p_{3/2} \otimes 0_{1}^{+}$   &0.02&0.20\\
           & $3d_{5/2} \otimes 2_{1}^{+}$   &0.03&0.13      &            & $3d_{5/2} \otimes 1_{1}^{-}$   &0.00&0.13\\
           & $2g_{9/2} \otimes 4_{1}^{+}$   &0.05&0.17      &            & $2g_{9/2} \otimes 3_{1}^{-}$   &0.03&0.08\\
           & $1i_{11/2}\otimes 4_{1}^{+}$   &0.01&0.06      &            &                                &    &    \\
           & $1i_{11/2}\otimes 6_{1}^{+}$   &0.03&0.09      &            &                                &    &    \\
           & $1j_{15/2}\otimes 7_{1}^{-}$   &0.01&0.09      &            &                                &    &    \\
\hline
$5/2^{+}$  & $3d_{5/2} \otimes 0_{1}^{+}$   &0.01&0.07      & $5/2^{-}$  & $3p_{3/2} \otimes 2_{1}^{+}$   &0.02&0.22\\
           & $3p_{3/2} \otimes 1_{1}^{-}$   &0.01&0.06      &            & $3d_{5/2} \otimes 3_{1}^{-}$   &0.01&0.11\\
           & $2g_{9/2} \otimes 2_{1}^{+}$   &0.05&0.13      &            & $2g_{9/2} \otimes 3_{1}^{-}$   &0.05&0.07\\
           & $1j_{15/2}\otimes 5_{1}^{-}$   &0.02&0.07      &            & $2g_{9/2} \otimes 5_{1}^{-}$   &0.00&0.05\\
           & $1i_{11/2}\otimes 6_{1}^{+}$   &0.02&0.17      &            &                                &    &    \\
\hline
$7/2^{+}$  & $3d_{5/2} \otimes 2_{1}^{+}$   &0.02&0.06      & $7/2^{-}$  & $3d_{5/2} \otimes 1_{1}^{-}$   &0.00&0.12\\
           & $2g_{9/2} \otimes 2_{1}^{+}$   &0.07&0.07      &            & $2g_{9/2} \otimes 1_{1}^{-}$   &0.05&0.09\\
           & $3p_{3/2} \otimes 3_{1}^{-}$   &0.01&0.09      &            & $3p_{3/2} \otimes 2_{1}^{+}$   &0.01&0.18\\
           & $3d_{5/2} \otimes 4_{1}^{+}$   &0.00&0.05      &            &                                &    &    \\
           & $2g_{9/2} \otimes 4_{1}^{+}$   &0.02&0.08      &            &                                &    &    \\
           & $1j_{15/2}\otimes 5_{1}^{-}$   &0.04&0.06      &            &                                &    &    \\
           & $1i_{11/2}\otimes 8_{1}^{+}$   &0.01&0.10      &            &                                &    &    \\
\hline
$9/2^{+}$  & $2g_{9/2} \otimes 0_{1}^{+}$   &0.06&0.10      & $9/2^{-}$  & $3d_{5/2} \otimes 3_{1}^{-}$   &0.01&0.08\\
           & $3d_{5/2} \otimes 2_{1}^{+}$   &0.01&0.12      &            & $3p_{3/2} \otimes 4_{1}^{+}$   &0.02&0.19\\
           & $2g_{9/2} \otimes 2_{1}^{+}$   &0.02&0.06      &            & $3d_{5/2} \otimes 5_{1}^{-}$   &0.00&0.07\\
           & $3p_{3/2} \otimes 3_{1}^{-}$   &0.01&0.08      &            & $2g_{9/2} \otimes 5_{1}^{-}$   &0.01&0.05\\
           & $1j_{15/2}\otimes 3_{1}^{-}$   &0.05&0.07      &            &                                &    &    \\
\hline
$11/2^{+}$ & $2g_{9/2} \otimes 2_{1}^{+}$   &0.06&0.04      & $11/2^{-}$ & $2g_{9/2} \otimes 1_{1}^{-}$   &0.05&0.08\\
           & $3d_{5/2} \otimes 4_{1}^{+}$   &0.01&0.07      &            & $1j_{15/2}\otimes 2_{1}^{+}$   &0.06&0.03\\
           & $2g_{9/2} \otimes 4_{1}^{+}$   &0.05&0.07      &            & $3d_{5/2} \otimes 3_{1}^{-}$   &0.00&0.15\\
           & $3p_{3/2} \otimes 5_{1}^{-}$   &0.01&0.08      &            & $3p_{3/2} \otimes 4_{1}^{+}$   &0.01&0.17\\
           & $1i_{11/2}\otimes 10_{1}^{+}$  &0.00&0.08      &            &                                &    &    \\
\hline
$13/2^{+}$ & $1j_{15/2}\otimes 1_{1}^{-}$   &0.05&0.05      & $13/2^{-}$ & $1i_{11/2}\otimes 1_{1}^{-}$   &0.07&0.10\\
           & $2g_{9/2} \otimes 2_{1}^{+}$   &0.08&0.14      &            & $3d_{5/2} \otimes 5_{1}^{-}$   &0.01&0.05\\
           & $3d_{5/2} \otimes 4_{1}^{+}$   &0.00&0.14      &            & $3p_{3/2} \otimes 6_{1}^{+}$   &0.01&0.13\\
           & $3p_{3/2} \otimes 5_{1}^{-}$   &0.01&0.09      &            & $3d_{5/2} \otimes 7_{1}^{-}$   &0.00&0.05\\
\end{tabular}
\end{ruledtabular}
\end{table*}

\begin{table*}[htbp]
\caption{\label{tab:E1E2} Theoretical  intraband and interband $E2$ and interband $E1$ transition rates (in Weisskopf units) for low-lying bands in $^{223, 225, 227}$Ra, in comparison with available experimental values \cite{NNDC}.}
\begin{ruledtabular}
\begin{tabular}{cccc}
\textrm{}&
\textrm{}&
\textrm{Theory}&
\textrm{Experiment}\\
\colrule
\footnotesize \centering {$^{223}$Ra}
&$B(E1;3/2_{band2}^{-}\rightarrow 5/2_{band1}^{+})$&0.00057&0.00050(9)\\
&$B(E1;7/2_{band2}^{-}\rightarrow5/2_{band1}^{+})$&0.00057&7.9$\times$$10^{-5}$(24)\\
&$B(E1;3/2_{band6}^{-}\rightarrow 1/2_{band5}^{+})$&0.00082&0.000160(20)\\
&$B(E1;3/2_{band6}^{-}\rightarrow 5/2_{band5}^{+})$&0&$5.2\times$$10^{-6}$(6)\\
&$B(E2;7/2_{band1}^{+}\rightarrow 5/2_{band1}^{+})$&86.59&70\\

&$B(E2;11/2_{band1}^{+}\rightarrow 7/2_{band1}^{+})$&74.80&$2.8\times$$10^{2}$(12)\\
&$B(E2;7/2_{band1}^{+}\rightarrow 3/2_{band1}^{+})$&31.48&44\\
&$B(E2;7/2_{band2}^{-}\rightarrow 5/2_{band2}^{-})$&107.90&$1.1\times$$10^{2}$(5)\\
&$B(E2;7/2_{band2}^{-}\rightarrow 3/2_{band2}^{-})$&60.56&10(6)\\
&$B(E2;3/2_{band6}^{-}\rightarrow 5/2_{band2}^{-})$&0.042&0.91(15)\\
&$B(E2;3/2_{band6}^{-}\rightarrow 7/2_{band2}^{-})$&0.38&1.65(24)\\
\hline
\footnotesize \centering {$^{225}$Ra}
&$B(E1;3/2_{band2}^{-}\rightarrow 1/2_{band1}^{+})$&0.0024&0.00081(20)\\
&$B(E2;3/2_{band1}^{+}\rightarrow 1/2_{band1}^{+})$&55.89&$\textgreater 2.0$\\
&$B(E2;5/2_{band1}^{+}\rightarrow 1/2_{band1}^{+})$&91.54&102(6)\\
\hline
\footnotesize \centering {$^{227}$Ra}
&$B(E1;3/2_{band2}^{-}\rightarrow 5/2_{band1}^{+})$&0.0037&0.00060(9)\\
&$B(E1;5/2_{band2}^{-}\rightarrow7/2_{band1}^{+})$&0.0058&0.0010(3)\\
&$B(E1;5/2_{band2}^{-}\rightarrow 3/2_{band1}^{+})$&0.0084&0.00030(7)\\
&$B(E1;3/2_{band4}^{-}\rightarrow 5/2_{band3}^{+})$&0.0012&$\textgreater 0.00067$\\
&$B(E1;3/2_{band4}^{-}\rightarrow 1/2_{band3}^{+})$&0.00093&$\textgreater 0.00050$\\
&$B(E1;1/2_{band4}^{-}\rightarrow 3/2_{band3}^{+})$&0.0023&$\textgreater 0.00045$\\
&$B(E1;1/2_{band4}^{-}\rightarrow 3/2_{band1}^{+})$&$8.3\times$$10^{-7}$&$\textgreater1.3$$\times$$10^{-6}$\\
&$B(E2;3/2_{band4}^{-}\rightarrow 5/2_{band2}^{-})$&0.14&$\textgreater 4.3$\\
\end{tabular}
\end{ruledtabular}
\end{table*}

\begin{table*}[htbp]
\caption{\label{tab:E3} Calculated $E3$ transition rates (in Weisskopf units) for low-lying energy levels from band 2 to the ground-state band in $^{223, 225, 227}$Ra.}
\begin{ruledtabular}
\begin{tabular}{ccccc}
\textrm{}&
\textrm{}&
\textrm{Th.}&
\textrm{}&
\textrm{Th.}\\
\colrule
\footnotesize \centering {$^{223}$Ra}
&$B(E3;9/2_{1}^{-}\rightarrow 3/2_{1}^{+})$&16.76     &$B(E3;11/2_{1}^{-}\rightarrow 5/2_{1}^{+})$&27.88\\
&$B(E3;13/2_{1}^{-}\rightarrow 7/2_{1}^{+})$&37.38    &$B(E3;15/2_{1}^{-}\rightarrow 9/2_{1}^{+})$&47.56\\
&$B(E3;17/2_{1}^{-}\rightarrow 11/2_{1}^{+})$&53.81   &$B(E3;19/2_{1}^{-}\rightarrow 13/2_{1}^{+})$&63.64\\
&$B(E3;21/2_{1}^{-}\rightarrow 15/2_{1}^{+})$&64.20\\
\hline
\footnotesize \centering {$^{225}$Ra}
&$B(E3;7/2_{1}^{-}\rightarrow 1/2_{1}^{+})$&38.26       &$B(E3;9/2_{1}^{-}\rightarrow 3/2_{1}^{+})$&29.27\\
&$B(E3;11/2_{1}^{-}\rightarrow 5/2_{1}^{+})$&71.21      &$B(E3;13/2_{1}^{-}\rightarrow 7/2_{1}^{+})$&60.78\\
&$B(E3;15/2_{1}^{-}\rightarrow 9/2_{1}^{+})$&83.75      &$B(E3;17/2_{1}^{-}\rightarrow 11/2_{1}^{+})$&79.58\\
&$B(E3;19/2_{1}^{-}\rightarrow 13/2_{1}^{+})$&90.93     &$B(E3;21/2_{1}^{-}\rightarrow 15/2_{1}^{+})$&97.79\\
\hline
\footnotesize \centering {$^{227}$Ra}
&$B(E3;9/2_{1}^{-}\rightarrow 3/2_{1}^{+})$&26.11     &$B(E3;11/2_{1}^{-}\rightarrow 5/2_{2}^{+})$&30.84\\
&$B(E3;13/2_{1}^{-}\rightarrow 7/2_{2}^{+})$&36.75    &$B(E3;15/2_{1}^{-}\rightarrow 9/2_{2}^{+})$&65.15\\
&$B(E3;17/2_{2}^{-}\rightarrow 11/2_{2}^{+})$&66.67   &$B(E3;19/2_{1}^{-}\rightarrow 13/2_{2}^{+})$&82.99\\
&$B(E3;21/2_{2}^{-}\rightarrow 15/2_{2}^{+})$&81.46\\
\end{tabular}
\end{ruledtabular}
\end{table*}

Tables \ref{tab:E1E2} and \ref{tab:E3} collect the results for intraband and interband electric quadrupole $E2$ transition rates and interband electric dipole $E1$ and octupole $E3$ transition rates for $^{223, 225, 227}$Ra. The overall agreement between theoretical predictions and experiment values is satisfactory, especially considering that the present calculation includes the full configuration space and, therefore, does not contain adjustable parameters in the form of effective charges. More specifically, most of the intraband $E2$ transition rates are reproduced very well, while the calculated interband $B(E2)$ values appear to be somewhat smaller than the experimental values. The theoretical $B(E1)$ values are generally large when compared to the data, but they are of the same order of magnitude. Some of these values are large, which is consistent with the pronounced octupole correlations predicted in these nuclei. Octupole correlations can be quantified by the calculated $E3$ transition rates in Table \ref{tab:E3}, where rather large $B(E3)$ values are predicted and they increase with angular momentum. In the Supplemental Material \cite{SupplementalMaterial} we also include model predictions for the intraband and interband $B(E2)$, and interband $B(E1)$, and $B(E3)$ values of $^{223, 225, 227}$Ra in Tables S1, S2, and S3, respectively.

\section{\label{secIV} Summary}

In summary, we have extended our microscopic CQC model to odd-mass nuclei characterized by both quadrupole and octupole shape deformations. The dynamics of the CQC Hamiltonian is determined by microscopically calculated single-nucleon and collective energies, quadrupole and octupole matrix elements, and pairing gaps corresponding to collective states of the even-mass core nuclei and spherical single-particle states of the odd nucleon. These are calculated using a quadrupole-octupole collective core Hamiltonian combined with a constrained relativistic Hartree-Bogoliubov model. The relativistic density functional PC-PK1 has been employed in the particle-hole channel, and a separable pairing force in the particle-particle channel. In the present version of the model the free parameters---the Fermi energy $\varepsilon_f$ and particle-vibration coupling strengths $\chi_2$ and $\chi_3$---are specifically adjusted to the experimental excitation energies of few lowest states. The model has been put to the test in a study of low-lying excitation spectra for the odd-A nuclei $^{223-227}$Ra. The theoretical results reproduce the available data on band structure, especially the parity doublets, intraband and interband $B(E2)$, and interband $B(E1)$. The near degeneracy of the parity doublets, pronounced $E1$ transitions, and large predicted $B(E3)$ values indicate the importance of pronounced octupole correlations in the low-energy structure of $^{223, 225, 227}$Ra.

\begin{acknowledgements}
This work was supported in part by the NSFC under Grants No. 11875225, No. 11790325, and No. 11475140,  the Inter-Governmental S\&T Cooperation Project between China and Croatia, and National Undergradruate Training Programs for Innovation and Entrepreneurship (201810635045). It was also supported in part by the QuantiXLie Centre of Excellence, a project cofinanced by the Croatian Government and European Union through the European Regional Development Fund - the Competitiveness and Cohesion Operational Programme (KK.01.1.1.01).
\end{acknowledgements}

\bibliography{CQCref}

\end{document}